\documentclass[letterpaper]{article} 
\usepackage{aaai2026}  
\usepackage{times}  
\usepackage{helvet}  
\usepackage{courier}  
\usepackage[hyphens]{url}  
\usepackage{graphicx} 
\usepackage{natbib}  
\usepackage{caption} 
\usepackage{algorithm}
\usepackage{algpseudocode}
\usepackage{algorithmicx}
\usepackage{amsmath}
\usepackage{amssymb}

\usepackage{newfloat}
\usepackage{listings}
\usepackage{tabularx}
\usepackage{array} 
\usepackage{multirow}
\usepackage{booktabs}
\newcolumntype{C}{>{\centering\arraybackslash}X}
\DeclareCaptionStyle{ruled}{labelfont=normalfont,labelsep=colon,strut=off} 
\floatstyle{ruled}
\newfloat{listing}{tb}{lst}{}
\floatname{listing}{Listing}
\nocopyright

\title{READ: Real-time and Efficient Asynchronous Diffusion for Audio-driven Talking Head Generation}
\author{
    Haotian Wang\textsuperscript{\rm 1},
    Yuzhe Weng\textsuperscript{\rm 1},
    Jun Du\textsuperscript{\rm 1}\thanks{Corresponding author},
    Haoran Xu\textsuperscript{\rm 2},
    Xiaoyan Wu\textsuperscript{\rm 2},
    Shan He\textsuperscript{\rm 2},
    Bing Yin\textsuperscript{\rm 2},
    Cong Liu\textsuperscript{\rm 2},
    Jianqing Gao\textsuperscript{\rm 2},
    Qingfeng Liu\textsuperscript{\rm {1,}}\textsuperscript{\rm 2}
}
\affiliations{
    \textsuperscript{\rm 1}University of Science and Technology of China, China \\
    \textsuperscript{\rm 2}iFLYTEK, China
}

\usepackage{bibentry}

\begin{document}

\maketitle

\begin{abstract}
The introduction of diffusion models has brought significant advances to the field of audio-driven talking head generation. However, the extremely slow inference speed severely limits the practical implementation of diffusion-based talking head generation models. In this study, we propose READ, a real-time diffusion-transformer-based talking head generation framework. Our approach first learns a spatiotemporal highly compressed video latent space via a temporal VAE, significantly reducing the token count to accelerate generation. To achieve better audio-visual alignment within this compressed latent space, a pre-trained Speech Autoencoder (SpeechAE) is proposed to generate temporally compressed speech latent codes corresponding to the video latent space. These latent representations are then modeled by a carefully designed Audio-to-Video Diffusion Transformer (A2V-DiT) backbone for efficient talking head synthesis. Furthermore, to ensure temporal consistency and accelerated inference in extended generation, we propose a novel asynchronous noise scheduler (ANS) for both the training and inference processes of our framework. The ANS leverages asynchronous add-noise and asynchronous motion-guided generation in the latent space, ensuring consistency in generated video clips. Experimental results demonstrate that READ outperforms state-of-the-art methods by generating competitive talking head videos with significantly reduced runtime, achieving an optimal balance between quality and speed while maintaining robust metric stability in long-time generation.
\end{abstract}

\begin{links}
    \link{Project Page}{https://readportrait.github.io/READ}
\end{links}

\section{Introduction}

Audio-driven talking head generation aims to generate videos of a person speaking an audio signal, which demonstrates significant value across multiple domains such as e-learning, film and game production, and human-computer interaction~\cite{thdsurvey1}. Evaluation criteria for audio-driven talking head generation models include the accuracy of lip synchronization with the input audio and the naturalism of the generated facial movements. In addition to these factors, the model's inference speed is also a crucial metric, as achieving real-time capabilities is essential for future human-computer interactive applications~\cite{thdsurvey2}.

Recently, the field of talking head generation has been greatly advanced by the introduction of diffusion models~\cite{diffusionsurvey}. Talking head generation frameworks built on the foundations of image or video diffusion models~\cite{emotivetalk} achieve more vivid performance than traditional methods~\cite{audio2head,sadtalker}. However, existing diffusion-based talking head generation models generally suffer from extremely slow inference speed, typically requiring tens to hundreds of seconds to generate a mere 5-second video~\cite{sonic,echomimic}, presenting a new challenge to this research field. The slow inference speed can be attributed to the following factors. First, the talking head generation task necessitates temporal alignment between speech features and video latents to ensure lip-sync accuracy. Existing methods typically employ a Variational Autoencoder (VAE)~\cite{vae} without temporal compression to achieve better alignment~\cite{sonic,echomimic,hallo}, yet increase the input token count and computational cost of the model. Second, conventional Denoising Diffusion Probabilistic Models (DDPM)~\cite{ddpm} or Denoising Diffusion Implicit Models (DDIM)~\cite{ddim} sampling methods require a large number of inference steps to generate high-fidelity video, substantially increasing inference time. Furthermore, considering extended generation, existing solutions mainly adopt overlap-and-fuse techniques~\cite{emotivetalk,sonic} or introduce an auxiliary network~\cite{hallo3,hallo} to maintain consistency between generated video clips, further increasing computational and time costs.

To address this challenge, in this research we introduce READ, the first end-to-end real-time diffusion-transformer-based audio-driven talking head generation framework. Our framework incorporates a temporal VAE with a high compression ratio of 32×32×8 pixels per token. To achieve better audio-visual alignment in the compressed latent space, we pre-train a Speech Autoencoder (SpeechAE) by self-supervising to generate temporally compressed speech latent codes corresponding to the compressed video latents. Then, an Audio-to-Video Diffusion Transformer (A2V-DiT) is designed to generate video latents under speech latent conditions efficiently. The training and inference processes of our framework are under the proposed Asynchronous Noise Scheduler (ANS), which implements an asynchronous add-noise forward process and an asynchronous motion-guided reverse process to effectively generate long-time videos.

\noindent In summary, our contributions are as follows: 
\begin{itemize}
    \item We propose an efficient Audio-to-Video Diffusion Transformer (A2V-DiT) model together with a pre-trained Speech Autoencoder (SpeechAE) to generate temporally aligned video latents under speech conditions at a relatively small runtime cost.
    \item We present an Asynchronous Noise Scheduler (ANS) for extended video diffusion, which achieves consistency between generated clips without extra computational cost.
    \item We further develop a real-time talking head generation framework that combines A2V-DiT and ANS, which can generate talking head videos at a 1:1 time ratio.
\end{itemize}

\section{Related Work}
\subsection{Audio-driven Talking Head Generation}
Audio-driven talking head generation aims to generate a talking person video conditioned on audio input, garnering increasing research interest due to its extensive application scenarios. Early research in audio-driven talking head generation primarily focused on achieving accurate lip synchronization with the input audio~\cite{wav2lip}. Subsequent works, such as Audio2Head~\cite{audio2head} and SadTalker~\cite{sadtalker}, advanced the field by producing more naturalistic head movements. More recent models, including DreamTalk~\cite{dreamtalk}, Diffused Heads~\cite{diffusedheads}, have further enhanced the expressiveness of the generated animations. Recently, a major shift occurred with the introduction of pretrained diffusion models~\cite{stablevideodiffusion,latentdiffusion}. Frameworks like Sonic~\cite{sonic}, EmotiveTalk~\cite{emotivetalk}, and Hallo~\cite{hallo} now leverage these powerful image or video diffusion priors to generate videos with improved fidelity and realism. However, a critical limitation of these models is their slow inference speed. We address this issue by proposing a novel framework specifically designed for fast talking head generation.

\subsection{Fast Diffusion Models}
Accelerating diffusion models is a major research focus~\cite{accelerationsurvey}. Progressive Distillation~\cite{progressivedistillation}, ADD~\cite{add}, LADD~\cite{ladd} and others~\cite{guideddiffusion,onestepdiffusion} focus on reducing diffusion steps. Ditto~\cite{ditto} and AniTalker~\cite{anitalker} employ motion-space diffusion to reduce tokens processed by the diffusion backbone for acceleration, yet face challenges with the naturalness of the generated video. In contrast, end-to-end video generation methods such as LTX-VIDEO~\cite{ltxvideo} and Wan~\cite{wan} utilize spatiotemporal compression in their VAEs to reduce computational cost. However, a critical issue arises when applying these VAEs to talking head generation, as temporal compression undermines the audio-visual alignment essential for accurate lip synchronization. To address this, we introduce a SpeechAE with self-supervised pre-training for synchronous speech feature compression to achieve better audio-visual alignment within end-to-end diffusion, and an Asynchronous Noise Scheduler (ANS) designed to ensure fast and stable extended inference.

\begin{figure*}[t]
\centering
\includegraphics[width=0.95\textwidth]{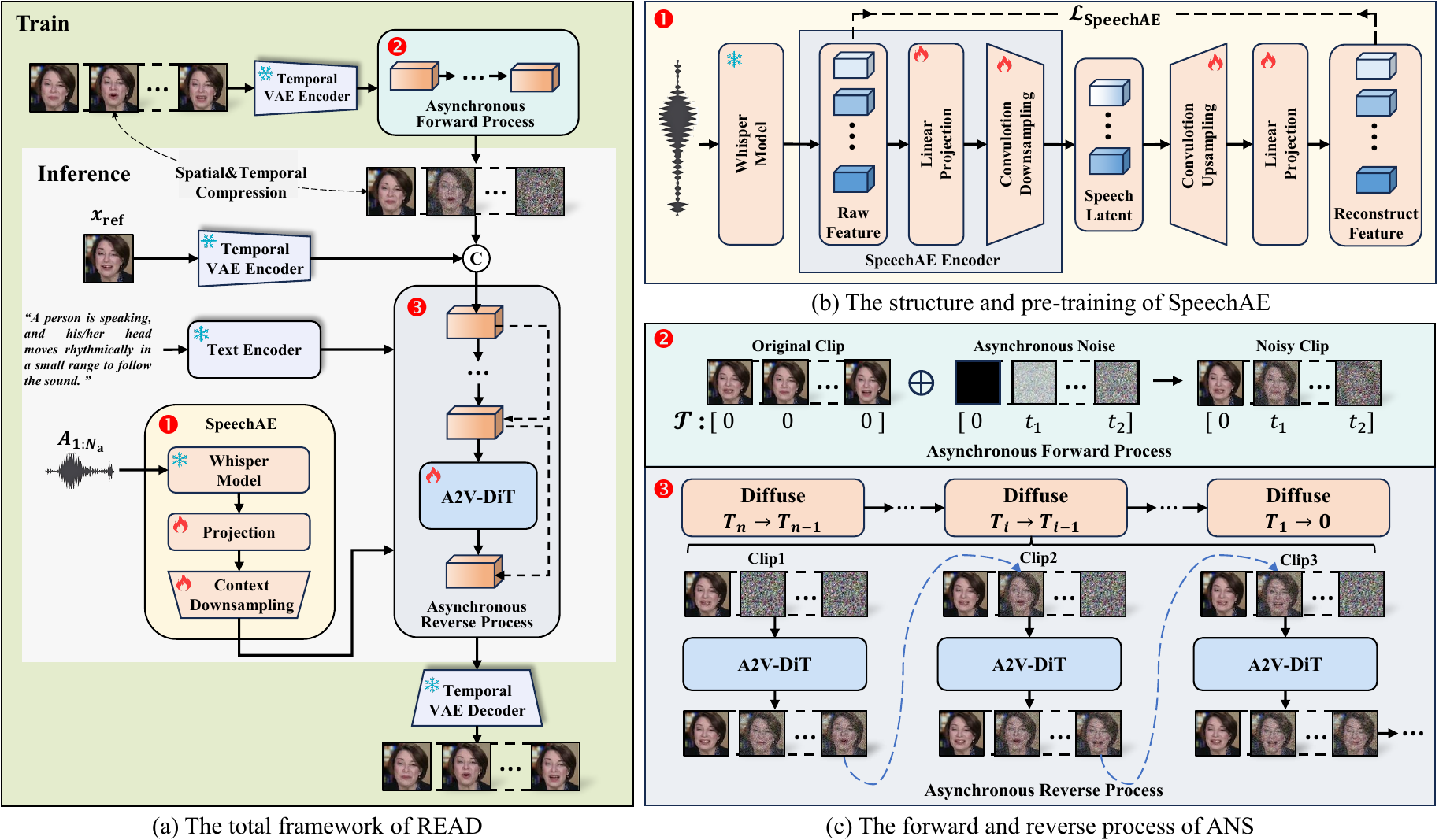} 
\caption{The framework of READ. During training, we first pre-train the SpeechAE for speech feature temporal compression, shown in (b). Then we train the total framework using the asynchronous forward process, shown in (c). During inference, we conduct the asynchronous motion-guided reverse process by ANS, also shown in (c).}
\label{fig:overall}
\end{figure*}

\section{Methods}
The total framework of READ is shown in Fig~\ref{fig:overall}. Sec.~\ref{sec:pre} outlines the necessary preliminaries relevant to our work. Sec.~\ref{sec:framework} details the proposed model architecture, including the pre-training procedure for the proposed Speech Autoencoder (SpeechAE). And the final section focuses on the training and inference methodology guided by our proposed Asynchronous Noise Scheduler (ANS).
\subsection{Preliminary}
\label{sec:pre}
\noindent\textbf{Task Definition}. Define the ground truth video sequence $ {\mathit{\boldsymbol{X}}}_{1:F} $. The audio-driven talking head generation takes a speech audio sequence $\mathit{\boldsymbol{A}_{1:F_{a}}} $ and a reference image $\mathit{\boldsymbol{I}_\text{ref}}$ as inputs. The output is the generated video $ \hat{\mathit{\boldsymbol{X}}}_{1:F} $ under only speech and reference image conditions.

\noindent\textbf{Flow Matching.}
Define $ {\mathit{\boldsymbol{Z}}}(0)$ as the original latents obtained by VAE, and $ {\mathit{\boldsymbol{Z}}}(t)$ as the noisy latents at timestep $t$. Flow Matching (FM)~\cite{flowmatching} is a generative method that leverages the principles of Ordinary Differential Equations (ODEs)~\cite{ode}. The central idea is to learn a continuous-time vector field ${\mathit{\boldsymbol{v}}}({\mathit{\boldsymbol{Z}}}(t), t)$ that transports samples from a simple noise distribution to the target data distribution $ {\mathit{\boldsymbol{Z}}}(0)$~\cite{flowmatching1,flowmatching2}:
\begin{equation}
    {{\mathit{{d}}}{\mathit{\boldsymbol{Z}}}(t)} = {\mathit{\boldsymbol{v}}}({\mathit{\boldsymbol{Z}}}(t), t){dt}
\end{equation}
The forward process of FM defines a probability path from the original distribution $ {\mathit{\boldsymbol{Z}}}(0)$ to $ {\mathit{\boldsymbol{Z}}}(t)$. The process can be formulated when using Gaussian probability paths to add synchronous Gaussian noise at timestep $t$ to $ {\mathit{\boldsymbol{Z}}}(0)$:
\begin{equation}
    {\mathit{\boldsymbol{Z}}}(t) = (1-t){\mathit{\boldsymbol{Z}}}(0) + t\boldsymbol{\epsilon},~\boldsymbol{\epsilon}\sim\mathcal{N}(0,\boldsymbol{I})
\end{equation}
The training objective of FM is for the model $\theta$ to learn the correct vector field ${\mathit{\boldsymbol{u}}}({\mathit{\boldsymbol{Z}}}(t), t)$, as follows:
\begin{equation}
    \mathcal{L}_{\text{FM}}(\theta) = \mathbb{E}_{t,Z(t)\sim p_{t}}||{\mathit{\boldsymbol{v}}}({\mathit{\boldsymbol{Z}}}(t), t) - {\mathit{\boldsymbol{u}}}({\mathit{\boldsymbol{Z}}}(t), t)||^2
\end{equation}
where ${\mathit{\boldsymbol{u}}}({\mathit{\boldsymbol{Z}}}(t), t)$ is the target gound-truth corresponding vector field. Our proposed ANS scheduler incorporates concepts from FM and introduces key innovations to both the forward (add-noise) and reverse (denoise) processes to guide the training and inference of our diffusion network.

\subsection{Fast Audio-to-Video Generation Framework}
\label{sec:framework}
In this section, we detail the overall architecture of READ, which is designed for efficient talking head generation. Shown in Fig.~\ref{fig:overall}, the READ framework consists mainly of three parts: Temporal VAE, SpeechAE, and A2V-DiT.  

\noindent\textbf{Temporal VAE for Video Compression.} 
Training and inference time for DiT models is dominated by the number of input tokens~\cite{dit}. To reduce the number of tokens processed by the backbone network to accelerate generation speed, we employ a temporal VAE with a high spatiotemporal compression ratio of 32×32×8 pixels per token from LTX-VIDEO~\cite{ltxvideo}. The principle can be formulated as follows:
\begin{equation}
    {\mathit{\boldsymbol{Z}}}(0) = \mathcal{E}_{\text{V}}({\mathit{\boldsymbol{X}}}(0)),~{\hat{\mathit{\boldsymbol{X}}}}(0) = \mathcal{D}_{\text{V}}({\mathit{\boldsymbol{Z}}}(0))
    \label{eq:vae}
\end{equation}
where ${\mathit{\boldsymbol{X}}}(0)\in\mathbb{R}^{H\times W\times F \times D_\text{v}}$ represents the video sequence, and ${\mathit{\boldsymbol{Z}}}(0)\in\mathbb{R}^{h\times w\times f \times d_\text{v}}$ are the compressed video latents. $\mathcal{E}_{\text{V}}$ and $\mathcal{D}_{\text{V}}$ denotes the encoder and decoder of VAE. 

\noindent\textbf{SpeechAE for Speech Feature Compression.} Unlike the text-to-video generation task, achieving precise temporal alignment between speech and video latents is particularly critical to achieve accurate lip synchronization in talking head generation. Although temporal VAE achieves a high compression ratio, it hinders audio-visual alignment because temporal compression disrupts the original correspondence between video and speech signals. We proposed SpeechAE self-supervised pre-training to perform synchronous temporal compression on raw speech features to address this limitation. Shown in (b) of Fig.~\ref{fig:overall}. Our SpeechAE integrates a frozen Whisper-tiny encoder~\cite{whisper} for speech feature extraction, as described below:
\begin{equation}
   \mathit{\boldsymbol{S}_{1:F}} = \mathcal{E}_{\text{Whisper}} (\mathit{\boldsymbol{A}_{1:F_{a}}} )
\end{equation}
The trainable part of SpeechAE also employs an encoder-decoder architecture, which consists of linear-based dimensionality transformation modules and temporal sampling modules built with 1D causal convolutional~\cite{cnn} layers, achieving the same temporal compression ratio as $\mathcal{E}_{\text{V}}$. The compressed speech latent codes $\mathit{\boldsymbol{C}\in\mathbb{R}^{f \times h_w\times d_{\text{A}}}}$ are generated by the SpeechAE encoder from $\mathit{\boldsymbol{S}}= [\mathit{\boldsymbol{s}}_1,..., \mathit{\boldsymbol{s}}_F] \in\mathbb{R}^{F\times H_w \times D_{\text{A}}}$, and reconstructed to $\hat{\mathit{\boldsymbol{S}}}=[\hat{\mathit{\boldsymbol{s}}}_1,..., \hat{\mathit{\boldsymbol{s}}}_F]$ through the decoder, as follows:
\begin{equation}
    {\mathit{\boldsymbol{C}}} = \mathcal{E}_{\text{A}}({\mathit{\boldsymbol{S}}}),~\hat{{\mathit{\boldsymbol{S}}}} = \mathcal{D}_{\text{A}}({\mathit{\boldsymbol{C}}})
\label{eqa:speechae}
\end{equation}
where $H_w$ and $h_w$ indicates the window sizes, $D_{\text{A}}$ and $d_{\text{A}}$ denotes the hidden dims. The quality of reconstruction can serve as a proxy for the information lost during compression~\cite{ae}. Effective reconstruction indicates that the latent codes $\mathit{\boldsymbol{C}}$ retain critical temporal information of source features ${\mathit{\boldsymbol{S}}}$. Based on this, we introduce a self-supervised pre-training phase on SpeechAE for the task of auto-encoding. To minimize the Euclidean distance between raw features ${\mathit{\boldsymbol{S}}}$ and reconstructed features $\hat{\mathit{\boldsymbol{S}}}$, we first apply a Mean Squared Error (MSE) loss, as follows:
\begin{equation}
    \mathcal{L}_\text{MSE}=||{\mathit{\boldsymbol{S}}}-\hat{\mathit{\boldsymbol{S}}}||^2
\end{equation}
Additionally, to enhance frame-level discrimination and preserve temporal variations of speech features, we introduce a contrastive loss to pull together speech features from corresponding frames while pushing apart features from distinct frames, as follows, with $\mathrm{sim}(\cdot)$ denotes cosine similarity:
\begin{equation}
  \mathcal{L}_{\text{CON}} =-\frac{1}{F} \sum_{i = 1}^{F} \mathrm{log} \left (\frac{\mathrm{exp}\left (\frac{\mathrm{sim}(\hat{\mathit{\boldsymbol{s}}}_i, {\mathit{\boldsymbol{s}}}_i)}{\tau}\right)}{\sum_{j = 1,j\neq i}^{F} \mathrm{exp}\left(\frac{\mathrm{sim}(\hat{\mathit{\boldsymbol{s}}}_i, {\mathit{\boldsymbol{s}}}_j)}{\tau}\right)}\right) 
\end{equation}

\noindent The final self-supervised loss function for SpeechAE pre-training is the combination of $\mathcal{L}_\text{MSE}$ and $\mathcal{L}_{\text{CON}}$, as follows:
\begin{equation}
    \mathcal{L}_{\text{SpeechAE}} = \alpha\mathcal{L}_\text{MSE} + \beta\mathcal{L}_{\text{CON}}
\end{equation}
Minimizing $\mathcal{L}_{\text{SpeechAE}}$ enables SpeechAE to produce temporally compressed speech latents that preserve the information in the raw speech features while aligning with the video latents, which serve as speech conditions to the A2V-DiT.

\noindent\textbf{A2V-DiT for Audio-driven Video Latents Generation.} To efficiently generate video latents ${\mathit{\boldsymbol{Z}}}(0)$ from speech latent codes $\mathit{\boldsymbol{C}}$, we introduce an A2V-DiT backbone. Each transformer block in A2V-DiT integrates three attention mechanisms: self-attention, 3D full-attention for text conditioning, and frame-level 2D cross-attention for audio conditioning. The self-attention module captures temporal dependencies across frames to enhance the consistency of the generated video latents. Since textual inputs describe the global video state, we apply 3D full-attention for text conditioning. Conversely, audio features demand precise temporal alignment with video latents. We leverage frame-level spatial cross-attention to generate video latents conditioned on the aligned speech latent codes ${\mathit{\boldsymbol{C}}}$ from SpeechAE, as formalized below:
\begin{equation}
  \mathit{\boldsymbol{H}}_{i}^{\text{A}} = \mathit{\boldsymbol{H}}_{i} +
  \rm{CrossAttn}(\mathit{\boldsymbol{H}}_{i},{\mathit{\boldsymbol{C}}})
  \label{eq:speechattn}
\end{equation}
where $\mathit{\boldsymbol{H}}_{i}, \mathit{\boldsymbol{H}}_{i}^{\text{A}}\in \mathbb{R}^{h\times w\times f \times d}$ denotes the unpatchified hidden states before and after frame-level audio cross-attention of the $i\!-\!\text{th}$ block of A2V-DiT. Our proposed design enables the efficient generation of video latents that are strictly synchronized with the corresponding speech conditions.

\subsection{Asynchronous Noise Scheduler (ANS)}
\label{sec:ANS}
The core concept of ANS leverages latent motion information during the lower-SNR stages of the diffusion process to guide motion generation in the higher-SNR stages, which maintains identity preservation while ensuring temporal consistency across extended generation sequences. The forward and reverse processes are detailed below.

        

        

        
\noindent\textbf{Asynchronous Forward Process for Training.} 
In contrast to traditional synchronous add-noise, our approach applies noise of different strengths to different positions of the video latents. Firstly, we define the first frame of the video latents as a motion frame to provide latent motion information to guide the motion generation of the following frames, while concatenating the reference frame to the front of the initial video latents to provide speaker identity, as follows:
\begin{equation}
    \mathit{\boldsymbol{z}}_\text{R}=\mathcal{E}_{\text{V}}(\mathit{\boldsymbol{I}_\text{ref}}), {\mathit{\boldsymbol{Z}}}(0)=[\mathit{\boldsymbol{z}}_\text{R},\mathit{\boldsymbol{z}}_1(0),...,\mathit{\boldsymbol{z}}_f(0)]
\end{equation}
Then we sample the asynchronous noise timestep $\boldsymbol{t}$ from the shifted-logit-normal distribution~\cite{sd3} based on the aforementioned latent structure that applies different noise timesteps to motion and reference frames, as follows:
\begin{equation}
    \boldsymbol{t}=[0,t_1,...,t_2],t_1,t_2 \!\sim \!\text{Sigmoid}(\mathcal{N}(\mu, \sigma)),t_1\!<\!t_2
\end{equation}
Next, the Gaussian noise $\boldsymbol{\epsilon}$ is added to the ${\mathit{\boldsymbol{Z}}}(0)$ based on the asynchronous timestep $\boldsymbol{t}$ to obtain the noisy latents ${\mathit{\boldsymbol{Z}}}(\boldsymbol{t})$, where $\boldsymbol{t}$ is broadcast to the dimensions of ${\mathit{\boldsymbol{Z}}}(0)$ beforehand.
\begin{align}
    \!{\mathit{\boldsymbol{Z}}}(\boldsymbol{t})\!&=\!(\boldsymbol{1}-\boldsymbol{t}) \odot {\mathit{\boldsymbol{Z}}}(0) + \boldsymbol{t} \odot \boldsymbol{\epsilon} \\ 
    \!\notag&= \![\mathit{\boldsymbol{z}_\text{R}},(1 \!-\!t_1)\mathit{\boldsymbol{z}}_1(0)\!+\!t_1\boldsymbol{\epsilon}, ...,(1\!-\!t_2)\mathit{\boldsymbol{z}}_f(0)\!+\!t_2\boldsymbol{\epsilon}]
\end{align}

\noindent The final training objective of the network parameters $\mathcal{S_{\theta}}$ is formulated as follows, conditioned on the audio latents $\boldsymbol{C}$:
\begin{align}
    \boldsymbol{v} &= \boldsymbol{\epsilon} - Z(0)\\
 \mathcal{L}_{\text{FM}} &= \mathbb{E}_{\boldsymbol{t},\boldsymbol{Z}(\boldsymbol{t})}||\boldsymbol{v} - \mathcal{S}_{\theta}(\boldsymbol{Z}(\boldsymbol{t}), \boldsymbol{C}, \mathit{\boldsymbol{z}_\text{R}},\boldsymbol{t})||^2
  \label{eq:denoise loss}
\end{align}
where $\boldsymbol{v}$ denotes the correct vector field under the Gaussian probability path of the asynchronous add-noise process.

\begin{algorithm}[t]
    \renewcommand{\algorithmicrequire}{\textbf{Input:}}
    \renewcommand{\algorithmicensure}{\textbf{Output:}}
    \caption{Asynchronous Forward Process}
    \label{power}
    \begin{algorithmic} 
        \Require 
            Time schedule $\{T_1,...,T_n\}$ ($T_n=0$),
            \\
            Reference image $\boldsymbol{I}_\text{ref}$ :
            $\mathit{\boldsymbol{z}}_\text{R}=\mathcal{E}_{\text{V}}(\mathit{\boldsymbol{I}_\text{ref}})$, 
            \\
            Speech latents $\boldsymbol{C} \in \mathbb{R}^{N \times h_w \times d_a}$,
            \\
            Noise vectors $\boldsymbol{\epsilon} = \{ \boldsymbol{\epsilon}_1,\!...,\!\boldsymbol{\epsilon}_N \}\!\in\!\mathcal{N}(\mathbf{0}, \mathbf{I})$, $\boldsymbol{Z}(T_1)=\boldsymbol{\epsilon}$ 
            
        \Ensure 
            Generated latents $\boldsymbol{Z}(0) \in \mathbb{R}^{h \times w \times N \times d_v}$
    
        \For{$i = 1$ \textbf{to} $n-1$} \Comment{Iterate over time steps}
        \State $\boldsymbol{Z}(T_i) \gets \{\boldsymbol{Z}_1(T_i),...,\boldsymbol{Z}_k(T_i)\}$ \Comment{Segment clips}
        \State $\boldsymbol{Z}_j(T_i)\! \gets\! \{\boldsymbol{z}_\text{R},\! \boldsymbol{z}_{1\!+\!(j-1)(f-1)}(T_i),\dots,\!\boldsymbol{z}_{1\!+\!j(f-1)}(T_i)\}$
        \State $\boldsymbol{C} \gets \{\boldsymbol{C}_1,\cdots,\boldsymbol{C}_k\}$
            \For{$j = 1$ \textbf{to} $k$} \Comment{Process each clip}
                \If{$j = 1$} \Comment{First clip}
                    \State $\boldsymbol{t}_j \gets [0, T_{i}, T_{i}, \cdots, T_{i}]$
                \Else \ \Comment{Subsequent clips}
                    \State $\boldsymbol{t}_j \gets [0, T_{i+1}, T_{i}, \cdots, T_{i}]$
                    \State $\boldsymbol{Z}_j(T_{i})[1] \gets \boldsymbol{Z}_{j-1}(T_{i+1})[f]$ \Comment{Guided}
                \EndIf
                \State $\boldsymbol{Z}_j(T_{i+1}) \xleftarrow[\text{FM}]{\text{CFG}} \mathcal{S}_{\theta}\big(\boldsymbol{Z}_j(T_{i}), \boldsymbol{C}_j, \boldsymbol{t}_j\big)$ \Comment{Generate}
            \EndFor
            \State $\boldsymbol{Z}(T_{i+1}) \gets \{\boldsymbol{Z}_1(T_{i+1}),...,\boldsymbol{Z}_k(T_{i+1})\}$ \Comment{Update}
        \EndFor
        
        \State \Return $\boldsymbol{Z}(0) \gets \{\boldsymbol{z}_\text{R}, \boldsymbol{z}_{1}(0),...,\boldsymbol{z}_{N}(0)\}$
    \end{algorithmic}
\end{algorithm}

\noindent\textbf{Asynchronous Reverse Process for Inference.} 
Following the training phase described above, the model learns the ability to leverage less-noisy motion frames to guide subsequent target frames generation. Our reverse sampling schedule leverages this mechanism to ensure long-term consistency during extended inference. As detailed in Algorithm 1, we first divide the target latent sequence of length $N$ into $k$ overlapping clips of length $f$ with one-frame overlap. The reference latent $\mathit{\boldsymbol{z}_\text{R}}$ is concatenated before each clip. The inference procedure employs a dual-loop architecture: In the outer loop, each clip is processed sequentially at the current timestep $T_{i}$. Notably, the initial clip undergoes free-form inference at noise timestep $\boldsymbol{t}=[0,T_{i},\cdots,T_{i}]$ without motion guidance due to the absence of preceding frames. For subsequent clips, we substitute each segment's first frame with the final frame from the previously generated clip, performing motion-guided inference at noise timestep $\boldsymbol{t}=[0,T_{i+1},\cdots,T_{i}]$. To balance runtime and performance during inference, we introduce two forms of Classifier-Free Guidance (CFG), Joint-CFG and Split-CFG. Joint-CFG conditions on the reference and speech conditions in a unified manner, as formulated below, where $\hat{\boldsymbol{v}}_{j}$ denotes the generated optical flow of $j\!-\!\text{th}$ clip:
\begin{equation}
    \hat{\boldsymbol{v}}_{j} = (1\!-\!\alpha)\mathcal{S}_{\theta}(\boldsymbol{Z}_{j}(\boldsymbol{t}), \emptyset,\boldsymbol{t})\!+\!\alpha \mathcal{S}_{\theta}(\boldsymbol{Z}_{j}(\boldsymbol{t}), \!\boldsymbol{C}_{j}, \!\mathit{\boldsymbol{z}_{R}},\!\boldsymbol{t})
\end{equation}
\noindent while Split-CFG applies CFG to each signal independently:
\begin{align}
\hat{\boldsymbol{v}}_{j} &= (1\!-\!\alpha\!-\!\beta)\mathcal{S}_{\theta}(\boldsymbol{Z}_{j}(\boldsymbol{t}), \emptyset,\boldsymbol{t})
\notag
\\&+\alpha \mathcal{S}_{\theta}(\boldsymbol{Z}_{j}(\boldsymbol{t}), \emptyset, \mathit{\boldsymbol{z}_{R}},\boldsymbol{t})
 + \beta \mathcal{S}_{\theta}(\boldsymbol{Z}_{j}(\boldsymbol{t}), \!\boldsymbol{C}_{j}, \!\mathit{\boldsymbol{z}_{R}},\!\boldsymbol{t})
\end{align}
\noindent The generated optical flow is then mapped back to the latent space via the FM scheduler~\cite{flowmatching}. This process iterates until all timesteps are processed, resulting in the generated temporally consistent latent sequence $\mathit{\boldsymbol{Z}}(0)$. 

\begin{table*}[t]
\centering
\begin{tabularx}{\linewidth}{@{}C|CC|C|C|C|C|C@{}}
\toprule
\textbf{Dataset} & \textbf{Method} & \textbf{Runtime(s) }& \textbf{FID ($\downarrow$)}  & \textbf{FVD ($\downarrow$)} & \textbf{Sync-C ($\uparrow$)} & \textbf{Sync-D ($\downarrow$)} & \textbf{E-FID ($\downarrow$)} \\ \midrule
\multicolumn{1}{c|}{\multirow{6}{*}{HDTF}} & FantasyTalking & 896.089 & 16.489 & 315.291 & 5.138 & 10.349 & 1.232 \\
\multicolumn{1}{c|}{} & Hallo & 212.002 & 15.929 & 315.904 & 6.995 & 7.819 & {\underline {0.931}} \\
\multicolumn{1}{c|}{} & EchoMimic & 124.105 & 18.384 & 557.809 & 5.852 & 9.052 & \textbf{0.927} \\
\multicolumn{1}{c|}{} & Sonic & 83.584 & 16.894 & {\underline {245.416}} & {\underline {8.525}} & \textbf{6.576} & 0.932 \\
\multicolumn{1}{c|}{} & AniPortrait & 76.778 & 17.603 & 503.622 & 3.555 & 10.830 & 2.323 \\
\multicolumn{1}{c|}{} & Ditto & 17.974 & {\underline {15.440}} & 399.965 & 5.458 & 9.565 & 2.659 \\
\multicolumn{1}{c|}{} & AniTalker & \underline{13.577} & 39.155 & 514.388 & 5.838 & 8.736 & 1.523 \\
\multicolumn{1}{c|}{} & Ours &  \textbf{4.421} & \textbf{15.073} & \textbf{235.319} & \textbf{8.658} & {\underline {6.890}} & 0.955 \\ \midrule
\multicolumn{1}{c|}{\multirow{6}{*}{MEAD}} & FantasyTalking & 896.089 & 46.617 & 257.077 & 4.536 & 10.699 & 1.510 \\
\multicolumn{1}{c|}{} & Hallo & 212.002 & 52.300 & 292.983 & 6.014 & 8.822 & {\underline {1.171}} \\
\multicolumn{1}{c|}{} & EchoMimic & 124.105 & 65.771 & 667.999 & 5.482 & 9.128 & 1.448 \\
\multicolumn{1}{c|}{} & Sonic & 83.854 & 47.070 & \textbf{218.308} & {\underline {7.501}} & \textbf{7.831} & 1.434 \\
\multicolumn{1}{c|}{} & AniPortrait & 76.778 & 54.621 & 531.663 & 1.189 & 13.013 & 1.669 \\
\multicolumn{1}{c|}{} & Ditto & 17.974 & \textbf{45.403} & 349.860 & 5.199 & 9.595 & 1.941 \\
\multicolumn{1}{c|}{} & AniTalker & \underline{13.577} & 95.131 & 621.528 & 6.638 & 8.184 & 1.553 \\
\multicolumn{1}{c|}{} & Ours &  \textbf{4.421} & {\underline {46.444}} & {\underline {224.738}} & \textbf{7.672} & {\underline {8.080}} & \textbf{1.043} \\ \bottomrule
\end{tabularx}
\caption{Overall comparisons on HDTF and MEAD. ``$\uparrow$'' indicates better performance with higher values, while ``$\downarrow$'' indicates better performance with lower values. The best results are \textbf{bold}, and the second-best results are \underline{underlined}.}
\label{tab:overall}
\end{table*}

\section{Experiments and Results}
\subsection{Experimental Setup} 
\label{sec:setup}
\noindent\textbf{Implementation Details}.
Experiments encompassing both training and inference are conducted on HDTF~\cite{hdtf} and MEAD~\cite{mead} datasets. 95\% data of both datasets is randomly allocated for training and the remaining 5\% for testing, ensuring no overlap between the partitions. We employ a two-stage training strategy. In the first stage, the SpeechAE is pre-trained with a learning rate of $1\times10^{-4}$. In the second stage, the entire audio-to-video backbone is trained at a resolution of $512\times512$ pixels and 121 frames, with a learning rate of $1\times10^{-5}$ and a batch size of 1. All reported results use 8-step sampling and Split-CFG with $\alpha=2.0$ and $\beta=6.0$ unless specified. The inference window length is set to match the training length with a motion overlap of one frame in latent space. Both training and evaluation are performed on NVIDIA A100 GPUs.

\noindent\textbf{Evaluation Metrics}. Generation performance is assessed using several metrics. For visual quality, we employ the Fréchet Inception Distance (FID)~\cite{fid} for image-level realism between synthesized videos and reference images and the Fréchet Video Distance (FVD)~\cite{fvd} for frame-level realism between synthesized and ground-truth videos; lower values indicate better performance for both metrics. Lip synchronization is measured with SyncNet~\cite{syncnet}, where a higher Synchronization Confidence (Sync-C) and a lower Synchronization Distance (Sync-D) indicate superior alignment with speech input. We further use the Expression-FID (E-FID) metric from EMO~\cite{emo} to measure the expression divergence between synthesized and ground-truth videos, with lower values indicating more faithful reproduction of expressions. Finally, we evaluate the efficiency of the diffusion backbone of each model by measuring the average runtime of the backbone per video (Runtime).

\noindent\textbf{Baselines}.
We benchmark our method against several SOTA open-source methods, including end-to-end diffusion methods such as Sonic~\cite{sonic}, EchoMimic~\cite{echomimic}, Hallo~\cite{hallo}, FantasyTalking~\cite{fantasytalking} and AniPortrait~\cite{aniportrait}, as well as motion-space diffusion methods like AniTalker~\cite{anitalker} and Ditto~\cite{ditto}. All comparisons are conducted on the same device using identical test data with the same length of 4.84s (121 frames) to ensure fair evaluation.

\subsection{Overall Evaluation}
As demonstrated in Tab.~\ref{tab:overall}, motion-space diffusion methods such as AniTalker and Ditto achieve reduced runtime compared to other end-to-end diffusion approaches. In contrast, our end-to-end approach achieves substantially lower latency in the backbone runtime than other methods, which represents a significant step forward for the acceleration of diffusion-based talking head generation. In addition to its speed, our approach also achieves highly competitive performance across all the evaluation metrics. On both HDTF and MEAD datasets, our model surpasses all competing methods in terms of Sync-C, while it also achieves SOTA or near-SOTA performance on FID, FVD, Sync-D, and E-FID. Notably, our model establishes superior E-FID performance on the emotion-rich MEAD dataset, validating its capability for generating expressive expressions faithful to the speech.

\subsection{Ablation Study} 
\label{sec:ablation}

\noindent\textbf{Effectiveness of SpeechAE}. 
To validate the contribution of our proposed SpeechAE module and self-supervised pretraining in maintaining audio-visual synchronization, we conduct an ablation study with three configurations:
\begin{itemize}
    \item \textbf{Full SpeechAE}: Method with pre-trained SpeechAE.
    \item \textbf{w/o Pre-training}: The SpeechAE module is trained from scratch with A2V-DiT without the pre-training stage.
    \item \textbf{w/o SpeechAE}: Speech features injected directly into A2V-DiT via linear projection without SpeechAE.
\end{itemize}
The experiment is carried out on the HDTF test set. Presented in Tab.~\ref{tab:ablation_speechae}, the results show that ablating only the pre-training stage leads to a noticeable degradation in lip-sync accuracy, with the Sync-C score decreasing by 0.693 and the Sync-D score increasing by 0.471. Furthermore, the complete removal of SpeechAE results in a substantial performance loss in lip-sync due to the temporal misalignment of raw audio features and the compressed video latents. These results validate the role of both the SpeechAE module and its self-supervised pre-training in achieving high-fidelity audio-visual synchronization for fast talking head generation.

\begin{table}[t]
\centering
\tabcolsep=0.036\linewidth
\begin{tabular}{@{}cccc@{}}
\toprule
\textbf{Configuration} & \textbf{FID ($\downarrow$)} & \textbf{Sync-C ($\uparrow$)} & \textbf{Sync-D ($\downarrow$)} \\ \midrule
Full SpeechAE & \textbf{15.073} & \textbf{8.658} & \textbf{6.890} \\
w/o Pre-training & 15.305 & 7.965 & 7.361 \\
w/o SpeechAE & 15.617 & 2.086 & 12.415 \\ \bottomrule
\end{tabular}
\caption{Ablation results of SpeechAE on HDTF dataset.}
\label{tab:ablation_speechae}
\end{table}

\begin{figure}[t]
    \centering
    \includegraphics[width=0.95\linewidth]{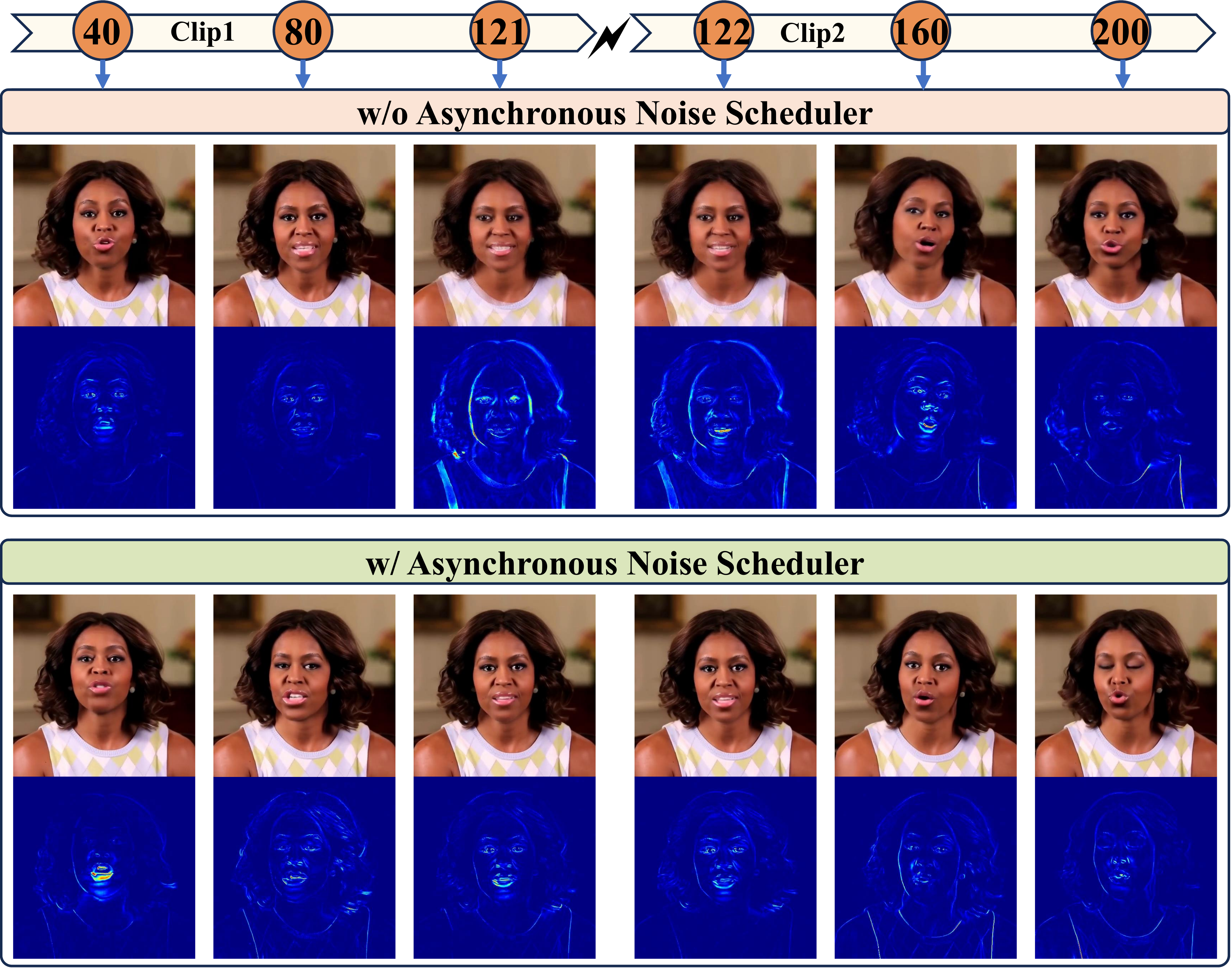}
    \caption{Ablation results of ANS on HDTF dataset.}
    \label{fig:ablation_ANS}
\end{figure}

\begin{table}[t]
\centering
\tabcolsep=0.0228\linewidth
\begin{tabular}{@{}ccccc@{}}
\toprule
\textbf{Frames} & \textbf{Duration(s)} & \textbf{FID($\downarrow$)}    & \textbf{Sync-C($\uparrow$)}  & \textbf{Sync-D($\downarrow$)}  \\ \midrule
121    & 4.840       & 15.073 & 8.658  & 6.891  \\
457    & 18.280      & 15.241 & 8.767  & 6.813  \\
1017   & 40.680      & 15.195 & 8.677  & 6.824  \\ \bottomrule
\end{tabular}
\caption{Results of different generation lengths on HDTF.}
\label{tab:ablation_longtime}
\end{table}

\begin{figure*}[t]
    \centering
    \includegraphics[width=1.0\linewidth]{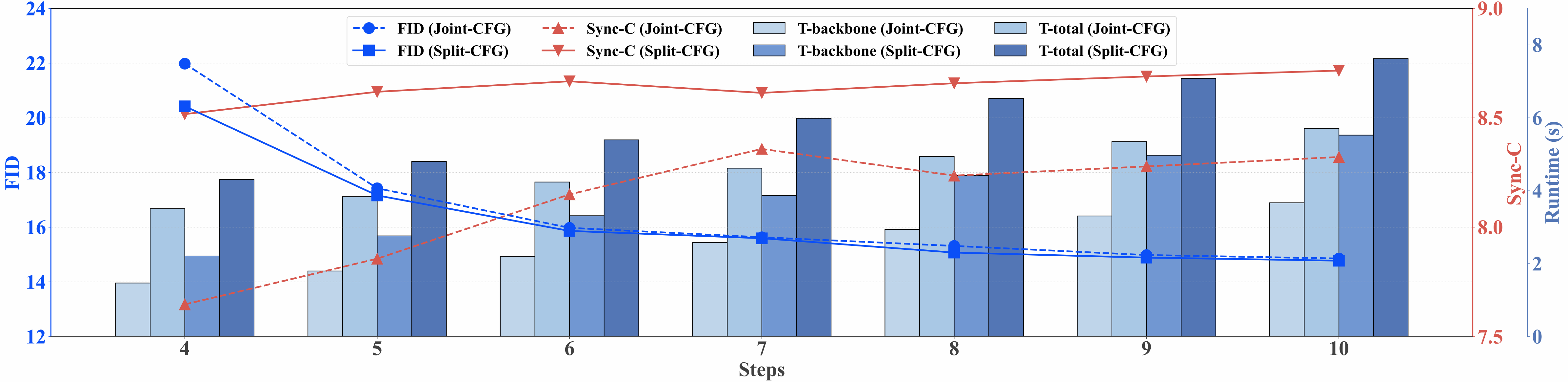}
    \caption{Trade-off between performance and runtime under different inference steps on HDTF dataset.}
    \label{fig:tradeoff}
\end{figure*}

\noindent\textbf{Effectiveness of Asynchronous Noise Schedule.}
To validate the contribution of our proposed Adaptive Noise Scheduler (ANS) to improving temporal consistency in generated videos, we conduct a qualitative ablation study by comparing the following two configurations:
\begin{itemize}
    \item Our Method with ANS \textbf{(w/ ANS)}: Utilizing the ANS Forward Scheduler for asynchronous add-noise during training and the ANS Reverse Schedule during inference.
    \item Baseline without ANS \textbf{(w/o ANS)}: Utilizing normal synchronous add-noise that adds the same strength of noise to the video latent during training and a standard clip concatenation strategy for inference~\cite{emotivetalk}.
\end{itemize}

\noindent For visualization, we generate two consecutive video clips, each 121 frames, and visualize the output by sampling one frame every 40 frames with special attention to the frames at the boundary of the two clips (frames 121 and 122). To assess the motion smoothness and consistency, we also visualize the difference heatmap between consecutive frames. Shown in Fig~\ref{fig:ablation_ANS}, results confirm that samples generated with our proposed ANS exhibit superior temporal consistency between video clips compared to the non-ANS baseline. While the baseline maintains reasonable intra-clip consistency with differences localized primarily to lip and face regions in the heatmap, it suffers significant inter-clip discontinuity, evidenced by pronounced error magnitudes spanning the entire talking head at the clip boundary (frames 121-122). In contrast, our method with ANS achieves consistent motion consistency both within and across clips, demonstrating smooth transitions between generated video clips. These results validate the important role of ANS in preserving temporal consistency for extended video generation.

\noindent\textbf{Effectiveness on Long-time Generation.} 
To test the performance stability of the generation quality and audio-visual synchronization of our framework with ANS on extended generation, we generated videos of varying lengths using the same model. Performance is assessed across different durations using three quantitative metrics: FID for visual quality, Sync-C, and Sync-D for lip-sync accuracy. Shown in Tab~\ref{tab:ablation_longtime}, the results validate that our model maintains consistent performance across varying generation lengths with no significant degradation in all three metrics, confirming the effectiveness of our framework and the proposed ANS in ensuring metric stability for extended video generation.

\noindent\textbf{Trade-off between Performance and Runtime.}
We conduct an ablation study to analyze the trade-off between inference quality and runtime of our framework. The study evaluates the distinct effects of Split-CFG and Joint-CFG strategies in the ANS reverse process and model performance under varying diffusion steps with two configurations:
\begin{itemize}
    \item \textbf{Split-CFG}: Using Split-CFG in ANS reverse process.
    \item \textbf{Joint-CFG}: Using Joint-CFG in ANS reverse process.
\end{itemize}
Each configuration is tested across a range of diffusion steps from 4 to 10. We measure generation quality and audio-visual synchronization while also recording the inference time of the backbone diffusion network and the total framework with VAE for each setting. The results are visualized in Fig.~\ref{fig:tradeoff}, which demonstrates a clear trade-off between runtime and performance. Reducing the number of inference steps proportionally decreases runtime but also degrades performance. This decline is more substantial for video quality (FID), particularly below 5 steps. Comparing the two CFG strategies, Split-CFG consistently outperforms Joint-CFG, especially on lip synchronization, albeit at the expense of increased runtime (almost by 50\%). This presents a clear trade-off that allows for the balancing of runtime with performance by choosing CFG strategies and inference steps.

\subsection{Case Study}
\begin{figure}[t]
    \centering
    \includegraphics[width=1.0\linewidth]{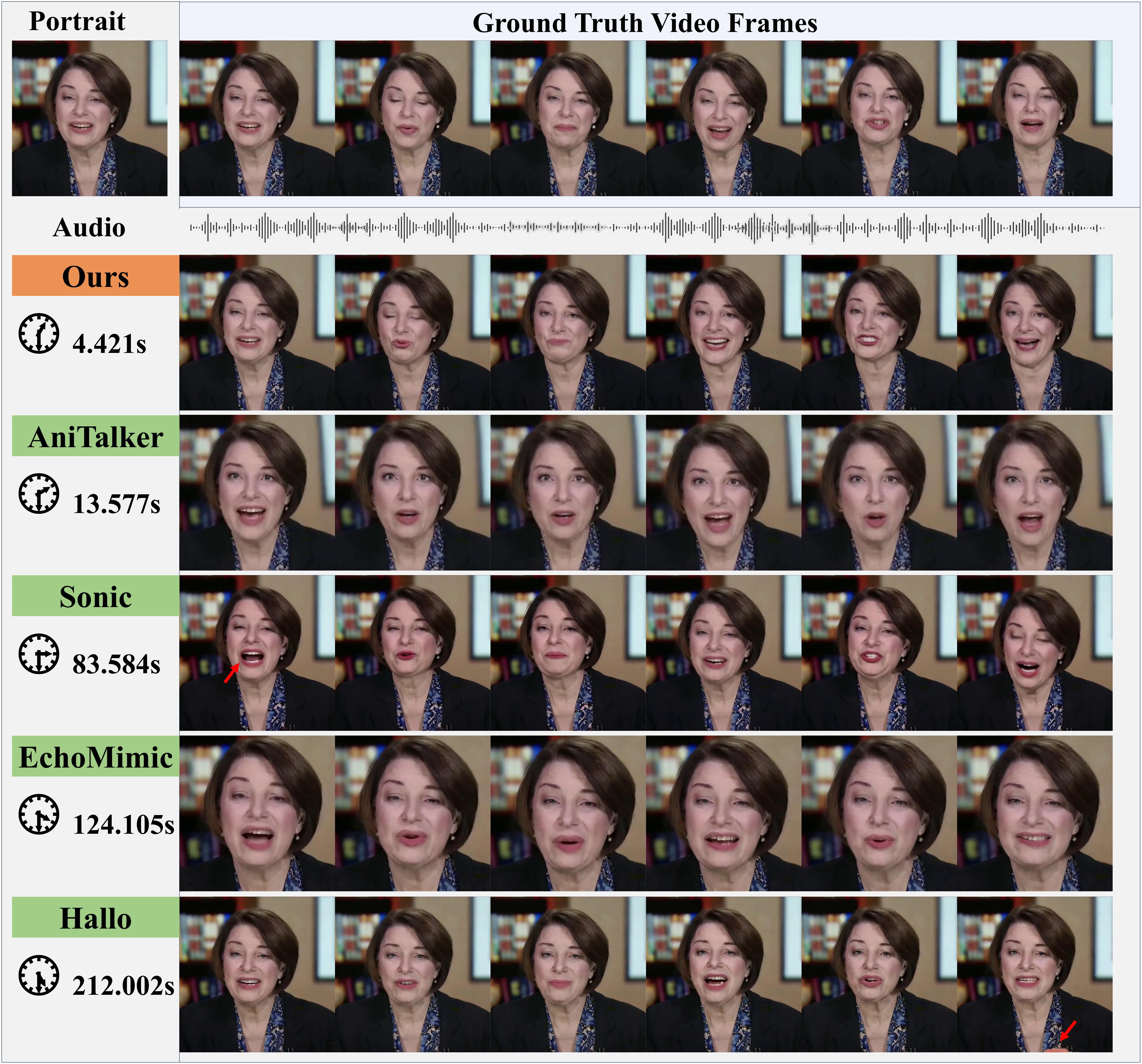}
    \caption{Case study of talking head generation methods.}
    \label{fig:casestudy}
\end{figure}

For a qualitative comparison of our model against other SOTA methods, we choose a representative case from the HDTF test dataset for detailed analysis. The frames are sampled at identical intervals from the videos generated by each model for visual comparison. The results are presented in Fig.~\ref{fig:casestudy}.
The visualization results demonstrate that AniTalker and EchoMimic require cropping or warping of the reference image, while failing to generate a video faithful to the reference image. Both methods also suffer from poor lip-sync accuracy, showing a mismatch of lip movements compared to the ground truth frames in the results. Sonic also presents inaccuracies in lip-sync at the start of the video clip. Meanwhile, Hallo suffers from generation instability, producing unexpected visual artifacts towards the end of the video (as indicated by the arrow). Compared to the results of other methods, the result generated by our method successfully maintains high fidelity to the reference image while achieving precise lip synchronization and high video quality, consistent with state-of-the-art performance.

\begin{table}[h]
\centering
\tabcolsep=0.0025\linewidth
\begin{tabular}{ccccc}
\hline
\textbf{Methods} & \textbf{Lip-Sync($\uparrow$)} & \textbf{Realness($\uparrow$)} & \textbf{Smooth($\uparrow$)} & \textbf{V-Qual($\uparrow$)} \\ \hline
Hallo & 3.303 & 2.786 & 2.714 & 2.819 \\
EchoMimic & 2.950 & 2.694 & 2.667 & 2.578 \\
Sonic & 4.111 & 3.756 & 3.875 & \textbf{3.994} \\
AniPortrait & 1.692 & 1.481 & 1.517 & 2.250 \\
AniTalker & 2.047 & 2.014 & 1.986 & 2.097 \\
Ours & \textbf{4.228} & \textbf{3.875} & \textbf{3.947} & 3.950 \\
GT & 4.656 & 4.528 & 4.597 & 4.578 \\ \hline
\end{tabular}
\caption{User study results of generation methods.}
\label{tab:usersstudy}
\end{table}

\subsection{User Study}
To qualitatively assess our model's performance, we conducted a user study involving 18 participants. We generated video samples based on 12 speech-image pairs using all 6 models. For each sample, the participants were required to give a rating (from 1 to 5, 5 is the best) on four aspects: (1) the lip sync quality (Lip-Sync), (2) the smoothness of generated motion (Smooth), (3) the realism of results (Realness), (4) the quality of video containing clarity and stability (V-Qual). Shown in Tab.~\ref{tab:usersstudy}, our method achieves the best results on Lip-Sync, Smooth, and Realness, and the second-best result on V-Q, highlighting its superior capabilities.

\section{Conclusion}
In this work, we propose READ, a novel DiT-based talking head generation framework that is able to generate real-time talking head videos. Our framework integrates a temporal VAE with a high compression ratio to reduce the token number of video latent processed by the network, thereby accelerating the generation speed. Specifically, a pre-trained SpeechAE module is proposed to generate temporally aligned speech latent codes corresponding to the video latent to achieve better audio-visual synchronization performance. Then we present a carefully designed A2V-DiT backbone to synthesize realistic talking head videos efficiently based on the speech latent codes generated by SpeechAE. Furthermore, we propose an ANS scheduler for both the training and inference of our entire framework, achieving asynchronous add-noise during training and asynchronous motion-guided inference during extended inference to generate temporally consistent long-time videos. Extensive experiments demonstrate the superiority of READ.

\section*{Acknowledgements}
This work was supported by the National Natural Science Foundation of China under Grant No. U25A20409.

\bibliography{main}

\appendix
\section{Dataset Details}
\label{sec:dataset}
\subsection{Dataset Introduction}
To effectively train our proposed model and ensure a comprehensive and rigorous evaluation, we utilize two of the most influential publicly available datasets in the field of talking head generation: the High-Definition Talking Face (HDTF)~\cite{hdtf} dataset and the Multi-view Emotional Audio-visual Dataset (MEAD)~\cite{mead}. The following subsection briefly highlights the primary advantages of both datasets and outlines the rationale for their selection in both model training and evaluation.

\noindent \textbf{High-Definition Talking Face (HDTF) Dataset.}
HDTF is a large-scale in-the-wild dataset celebrated for its high resolution and diversity. Collected from YouTube, the HDTF dataset contains a total of 15.8 hours of talking human videos, which features 362 distinct subjects speaking over 10,000 unique sentences, ensuring a rich variety of identities and speech content. We selected the HDTF dataset for our study due to several primary advantages of HDTF, which are detailed as follows:
\begin{itemize}
    \item High Fidelity: Most of the videos in HDTF are in 720p or 1080p, consisting of single-subject, frontal-view videos that exhibit both natural dynamics and synchronous high-quality lip movements, which provides an optimal basis for the training of our model to learn identity preservation and accurate lip synchronization in generated videos. 
    \item Real-World Variability: HDTF encompasses a wide range of ages, ethnicities, head poses, lighting conditions, and backgrounds, making it ideal for evaluating models' generalization to unseen scenarios.
\end{itemize}

\noindent \textbf{Multi-view Emotional Audio-visual Dataset (MEAD)} MEAD is a large-scale talking head video dataset specifically designed for research on emotional and multi-view talking head generation, which contains talking head videos of 60 actors. We selected the MEAD dataset for our study due to several primary advantages of MEAD, which are outlined below:
\begin{itemize}
    \item Rich Emotions: MEAD features 60 actors portraying 8 different emotions (e.g., happy, sad, angry) at 3 distinct intensity levels, which is suitable for evaluating models' ability to generate expressive talking head videos.
    \item Multi-View Setting: Videos of MEAD are captured from 7 different camera angles under a clean background. This provides clean and high-quality data for model training.
\end{itemize}

\subsection{Data Pre-processing} 
This section details the processing pipeline applied to the original HDTF and MEAD datasets to derive the training and test sets. Our data processing comprises three key steps: video processing, audio processing, and data filtering.

\noindent\textbf{Video Processing.} 
For the video modality, we perform a video pre-processing procedure with fps standardization and frame cropping. Firstly, we standardize the frame rate of all videos to 25 fps using the FFmpeg toolkit and extract the frames from videos. Subsequently, each video is cropped to a square aspect ratio (1:1). To achieve this, we first extract 68 facial landmark sequences from the extracted frames with the OpenFace~\cite{openface} toolkit. Based on the global maximum and minimum facial landmark coordinates of each video, we determined a fixed bounding box for cropping. Finally, to optimize storage efficiency, the cropped keyframes are re-encoded into a video sequence via FFmpeg, producing the final processed video data utilized for model training.

\noindent\textbf{Audio Processing.} For the audio track of each video, we first extract the audio signals from raw video data by the FFmpeg toolkit. Then, the audio signals are downsampled to 16 kHz and converted into a mono channel uniformly for speech feature extraction. Subsequently, we utilize the Whisper-tiny encoder~\cite{whisper} to extract speech features from the converted audio signals offline, which serve as the speech features input for model training. This process can be formulated as follows: 
\begin{equation}
   \mathit{\boldsymbol{S}_{1:N}} = \mathcal{E}_{\text{Whisper}} (\mathit{\boldsymbol{A}_{1:N_{a}}} )
\end{equation}
\noindent where $\mathit{\boldsymbol{A}_{1:N_{a}}}$ is the raw audio stream with a sample rate of 16 kHz and $\boldsymbol{S}_{1:N} \in \mathbb{R}^{N\times H_w \times D_{\text{A}}}$ denotes the offline speech feature, $N$ denotes the frames number of 25Hz, $H_w=10$ is the audio window size and $D_{\text{A}}=384$ is the hidden dim of the audio feature. This offline feature extraction process enhances the efficiency of the model training process.

\noindent\textbf{Data Filtering.} 
Subsequently, to ensure the purity and high quality of training data, we perform additional data filtering to retain only clean frontal speaker video data. This filtering stage comprises two key processes: hand filtering and subtitle content filtering. First, to mitigate the generation of visual artifacts caused by hand occlusions, we filter out videos containing significant hand movements. This is achieved by using the MediaPipe~\cite{mediapipe} toolkit for hand keypoint detection, with a preset hand detection confidence threshold of 0.8. The video clips detected with hands above the threshold are removed. Second, the presence of text, such as subtitles, is undesirable for the training of audio-driven talking head generation models. We address this by using the pre-trained Character Region Awareness for Text Detection (CRAFT)~\cite{craft} model to identify text regions within the video frames. Focusing specifically on the lower region of the frame where subtitles most frequently appear, we filter out any videos containing detected subtitles. This data filtering pipeline ensures that our final training dataset consists of high-quality frontal-view talking head data without hand occlusions and subtitles, providing a solid foundation for efficient and effective training of our model.

\section{Model Architecture Details of READ}
\label{sec:architecture}
In this section, we provide a detailed description of the architecture of our entire framework, including the detailed encoder-decoder structure of the SpeechAE module and the design of the A2V-DiT backbone.

\subsection{Detailed Architecture of SpeechAE.} 
\begin{figure}[t]
    \centering
    \includegraphics[width=1.0\linewidth]{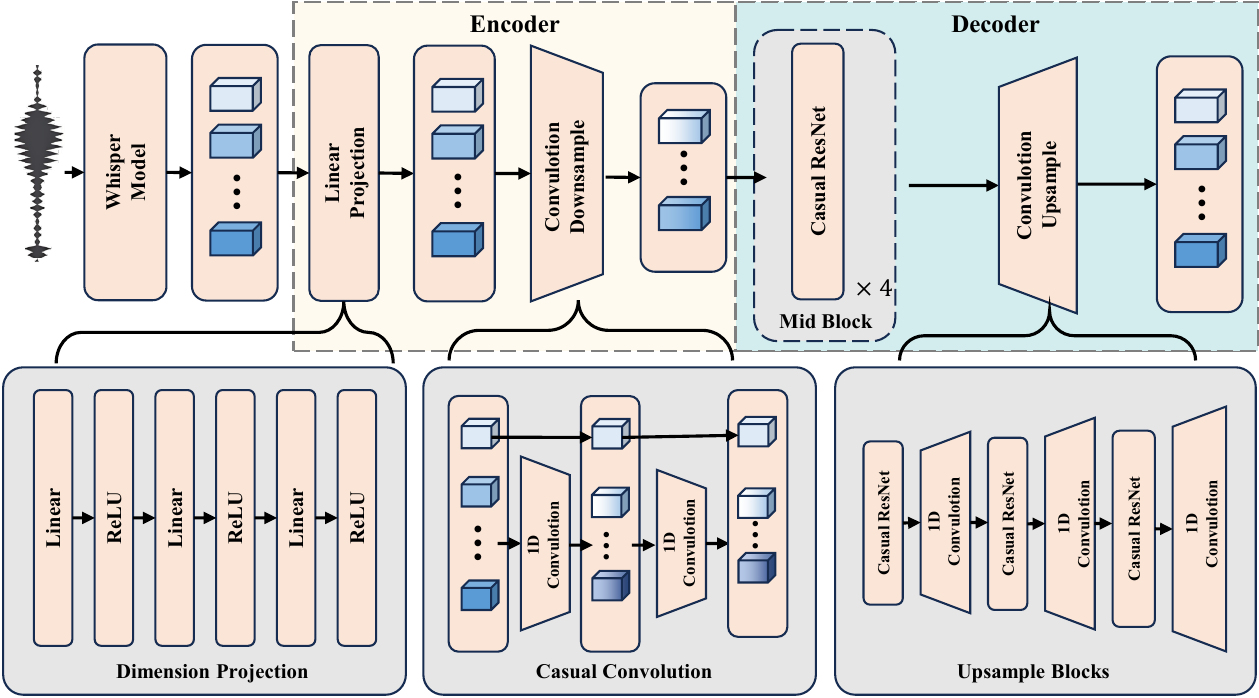}
    \caption{Detailed architecture of SpeechAE.}
    \label{fig:arch_speechae}
\end{figure}
The detailed architecture of our proposed SpeechAE is shown in Fig.~\ref{fig:arch_speechae}. As shown in Fig.~\ref{fig:arch_speechae}. The SpeechAE employs a fully convolutional encoder-decoder architecture. The encoder comprises two primary components: a linear projection block for feature dimension transformation, and three convolution downsampling blocks for the temporal compression of audio features. A key aspect of the design of SpeechAE is ensuring precise alignment with the video latent space. To facilitate this, the first frame of the raw audio features is processed separately, the same as the video latent representation obtained from our temporal VAE. The raw audio features $\boldsymbol{S}_{1:F} \in \mathbb{R}^{F\times H_w \times D_{\text{A}}}$ are processed by the SpeechAE Encoder to produce temporally compressed speech latent codes$\mathit{\boldsymbol{C}\in\mathbb{R}^{f \times h_w\times d_{\text{A}}}}$, where $h_w=2, d_{\text{A}}=2048$ and $f=(F-1)//8+1$ in our implementations, as formulated in Eq.~\ref{eqa:speechae} of Sec.~\ref{sec:framework}.

\subsection{Detailed Architecture of A2V-DiT.}
\begin{figure}[t]
    \centering
    \includegraphics[width=1.0\linewidth]{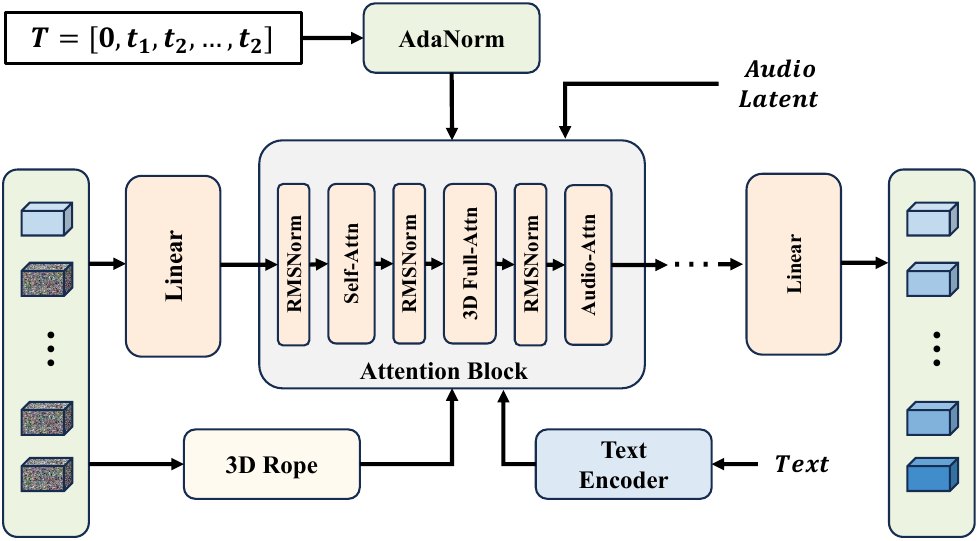}
    \caption{Detailed architecture of A2V-DiT.}
    \label{fig:a2v-dit}
\end{figure}

Our A2V-DiT model is built upon a Diffusion Transformer (DiT)~\cite{diffusiontransformer} architecture, as illustrated in Fig.~\ref{fig:a2v-dit}. The input to the backbone is the initial noisy video latent after being patchified. The shape of the patchified video latent is $[h\times w \times f, d]$, where $h=16,w=16,f=16,d=128$ in our implementation. The patchified video latent is first projected into a consistent latent space of hidden dimensionality 2048 via a linear layer. To encode positional information across both space and time, we apply 3D Rotary Position Embeddings (RoPE)~\cite{rope1,rope2} to the pachified video tokens. The core of the A2V-DiT model consists of 24 standard transformer blocks, as detailed in Sec.~\ref{sec:framework}. The model is conditioned on several inputs to guide the generation process:
\begin{itemize}
    \item \textbf{Speech Condition}: Processed by the previously described SpeechAE encoder, the resulting speech latent codes serve as input to a spatial audio attention module for audio conditioning, which is formulated in Eq.~\ref{eq:speechattn}.
    \item \textbf{Text Condition}: Text embeddings are first extracted from the text prompt and text attention mask by the T5~\cite{t5} encoder and then fed into a 3D full-attention module for text conditioning. Notably, for the text conditioning during training, we use a global text embedding that is pre-extracted offline using the T5 model. Further details are provided in Sec.~\ref{sec:trainingcfg}.
\end{itemize}
\noindent Additionally, the asynchronous noise timestep is encoded via an AdaLayerNorm~\cite{diffusiontransformer} module and incorporated into each attention layer to condition the noise prediction. Finally, the output of the transformer blocks is transformed to the target video latent. Our model design ensures effective audio-visual alignment while maintaining high computational efficiency, forming a foundation for fast talking head video generation.

\section{Training and Inference Details.}
\label{sec:trainvaldetails}
This section details the training and inference pipeline of the READ framework, including hyperparameter configurations. The elaboration consists of three primary parts: model initialization, SpeechAE pre-training, and training of the A2V-DiT backbone.

\subsection{Model Initialization.}
Our temporal VAE adopts the architecture and pre-trained weights from LTX-VIDEO~\cite{ltxvideo}. As formalized in Eq.~\ref{eq:vae}, the temporal VAE achieves a spatiotemporal compression ratio of 32×32×8 (spatial × spatial × temporal), yielding a latent space dimension of 128. For the main A2V-DiT backbone, both the self-attention modules and the 3D full-attention modules are initialized with the corresponding weights from the pre-trained weights of LTX-VIDEO~\cite{ltxvideo}. In contrast, our proposed spatial audio attention modules and the SpeechAE module for audio conditioning are initialized randomly.

\subsection{Devices Information.} 
All training and inference procedures are conducted on NVIDIA A100 GPUs, with the PyTorch library and CUDA version 12.6. During training, the model occupied approximately 46GB of GPU memory. For inference, generating a 4.8-second video clip required about 12GB of GPU memory. The total training process takes up approximately 240 GPU hours, including the pre-training of SpeechAE and the training of the A2V-DiT backbone.

\begin{figure*}[t]
    \centering
    \includegraphics[width=1.0\linewidth]{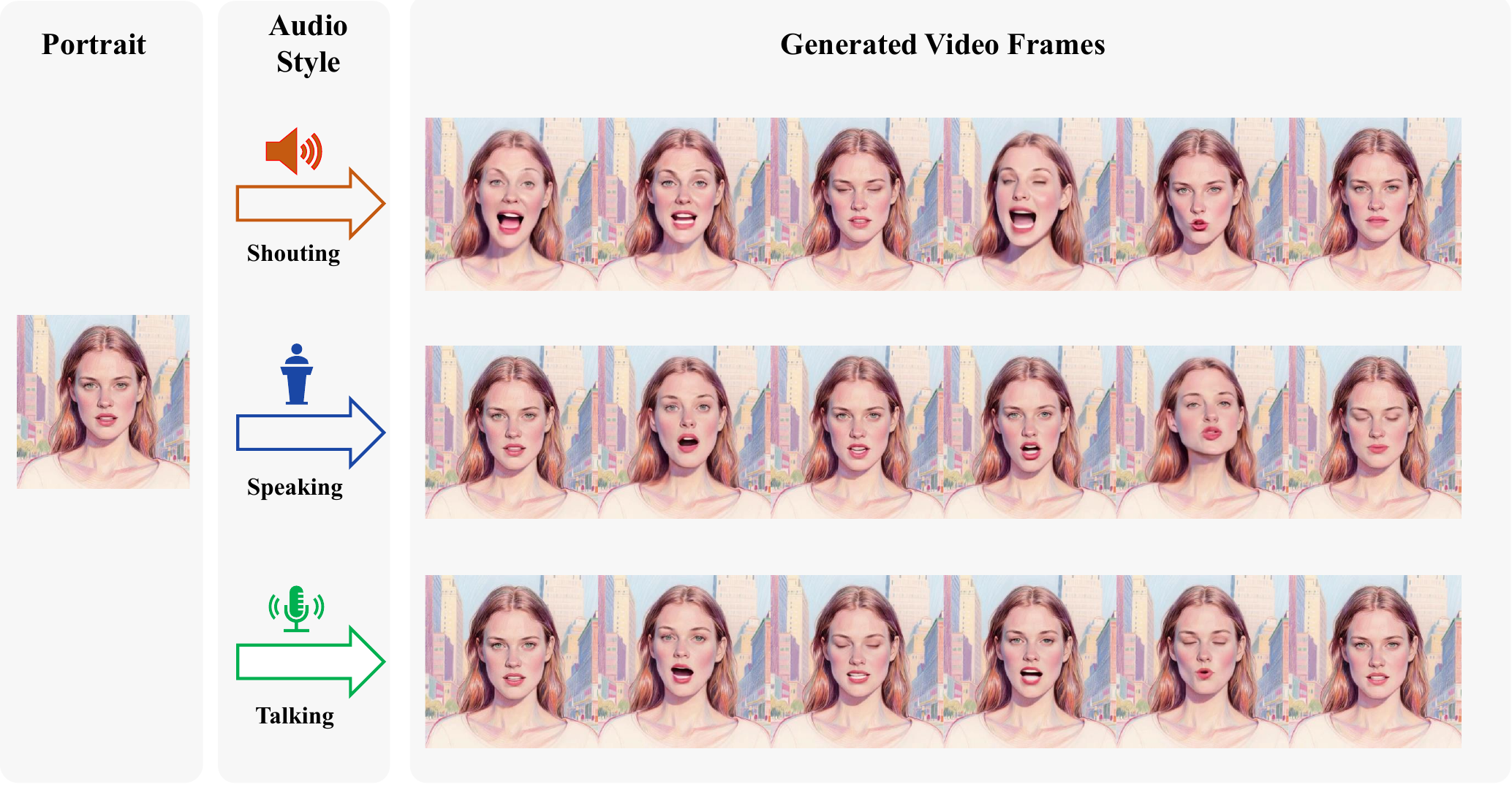}
    \caption{Generation results conditioned on the same reference image and different styles of audio input.}
    \label{fig:multistyle}
\end{figure*}

\subsection{Training Configurations.} 
\label{sec:trainingcfg}
As described in Sec.~\ref{sec:setup}, our training methodology consists of a two-stage process. We first pre-train the SpeechAE module with a learning rate of $1\times10^{-4}$. The pre-training is guided by the loss function defined in Eq.~\ref{sec:framework} of the main paper, where we set the balancing coefficients to $\alpha=10$ and $\beta=1$. Subsequently, we train the main A2V-DiT backbone with a lower learning rate of $1\times10^{-5}$, and the pre-trained SpeechAE module is jointly fine-tuned with the rest of the network. During the training process guided by ANS, we sample 
asynchronous noise from the shifted-logit-normal distribution~\cite{sd3}, as described in Sec.~\ref{sec:ANS} with $\mu=2.05$ and $\sigma=1.0$. We also introduce dropout training with three distinct dropout strategies during the training process guided by the proposed ANS:
\begin{itemize}
    \item \textbf{Identity Dropout}: With a probability of $p=0.1$, we randomly zeroize the reference image for enhancing the robustness of the model.
    \item \textbf{Audio Dropout}: The audio condition is randomly zeroized with a probability of $p=0.1$, reinforcing the effect of speech audio condition in audio-driven talking head generation. 
    \item \textbf{Motion Prior Dropout}: We introduce a motion prior dropout to the ANS forward noising process, with 50\% probability we apply asynchronous noise (lower-SNR noise for motion frames) to simulate conditional generation guided by a motion prior. For the remaining 50\%, we apply synchronous noise of uniform intensity across all frames, simulating an unconditional generation scenario.
\end{itemize}
For the text condition, we utilize a global text prompt for the description of talking head videos, which is ``\textit{A person is speaking, and his head moves rhythmically in a small range to follow the sound}''. To ensure more accurate gradient computation and reduce memory usage, we also incorporate a gradient accumulation strategy during training. We set the gradient accumulation steps to 4 and also apply the gradient checkpointing strategy~\cite{gradientcheckpointing} to ensure more accurate gradient computation and reduce memory usage.

\subsection{Inference Configurations.}
For inference, we generate video clips with a length of 121 frames, which is identical to the training sequence length. To ensure motion-guided generation by the ANS reverse process, we employ an overlap of 8 frames in the pixel space (corresponding to 1 frame in the latent space), which serves as a motion prior for the subsequent clip. As detailed in Section 1, we utilize two distinct Classifier-Free Guidance (CFG) strategies: Joint-CFG and Split-CFG. The guidance scales for these strategies are set as follows:
\begin{itemize}
    \item For Joint-CFG, we apply a unified guidance scale of $\alpha=2.0$ for both reference image and speech conditions.
    \item For Split-CFG, the guidance scale for the reference image is set to $\alpha=2.0$, while the scale for the speech audio condition is set to a stronger value of $\beta=6.0$.
\end{itemize}

\section{Additional Qualitative Results}
\label{sec:additionalresults}
In this section, we systematically evaluate the generation quality and stability of our proposed framework. To challenge the generalization capability of READ, we utilize inputs of varying speaker styles and speech content, and evaluate the quality of the generation results. Some generation results are visualized in Fig.~\ref{fig:multistyle} and Fig.~\ref{fig:multipotraits}. More results are included in the supplementary video. Please note that the resolution of the supplementary video has been reduced to comply with file size restrictions. 

\begin{figure*}[t]
    \centering
    \includegraphics[width=1.0\linewidth]{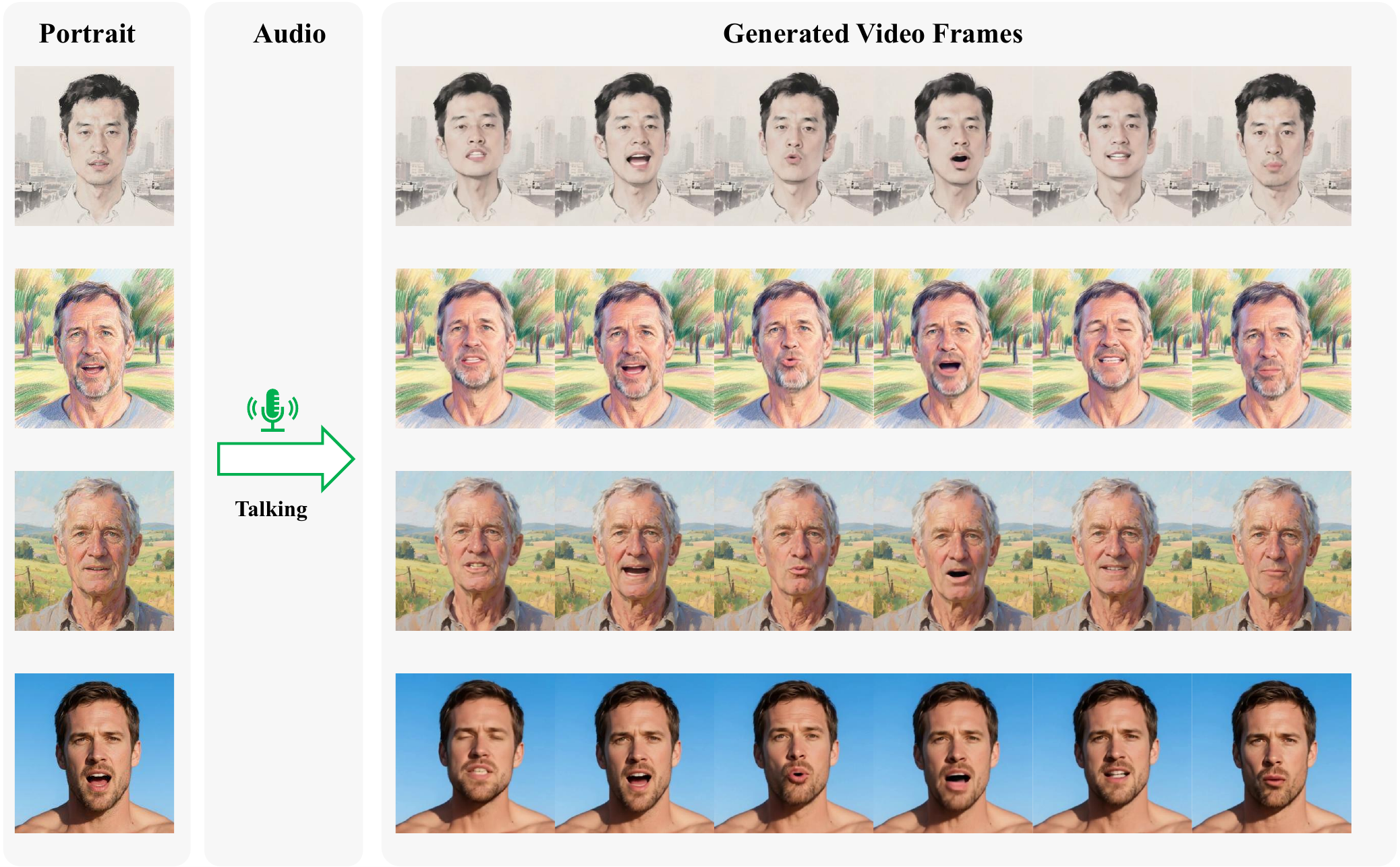}
    \caption{Generation results conditioned on the same talking audio and different styles of reference images.}
    \label{fig:multipotraits}
\end{figure*}

\subsection{Real-time and Expressive Generation} The supplementary video comprehensively demonstrates real-time generated talking head video results from READ. These results validate that READ can synthesize realistic, natural, and highly expressive talking head videos conditioned on speech and reference images, which demonstrates that our proposed method not only leverages the high quality and diversity inherent in full-video diffusion models but also effectively overcomes the challenges posed by the extremely slow inference speeds of diffusion-based talking head generation models. Consequently, READ achieves an optimal balance between generation performance and runtime, providing significant research and practical value for real-time audio-driven talking head generation applications.

\subsection{Performance on Multiple Audio Styles} 
To test our model's generalization performance on diverse audio styles, we used a single reference image paired with various audio clips as conditional inputs. Specifically, we tested three distinct audio styles: shouting, conversational speech, and talking. The corresponding output talking head videos generated by our model under these three distinct audio conditions are shown in Fig.~\ref{fig:multistyle}. 
As illustrated in Fig.~\ref{fig:multistyle}, our model generates distinct results when conditioned on audio inputs of varying styles. When the input is a shouting audio, the model produces a highly expressive performance, characterized by larger mouth movements, exaggerated facial expressions, and vigorous head movements, indicating a strong emotional state. For speech-style audio input, the generated talking head video exhibits rhythmic head motion and natural expressions, aligning with the speech. In contrast, when conditioned on conversational audio, head movements are more subtle compared to the shouting and speaking styles, consistent with the affective state of a typical conversation. These findings demonstrate that the READ can generate diverse talking head states that are consistent with the input audio, highlighting its superior capability in generating varied and realistic talking head videos.

\subsection{Performance on Multiple Image Styles}
To assess the generalization performance of our model across diverse reference image styles, we generated talking head videos using the same speech but conditioned on multiple reference styles unseen during training. The selected reference styles encompass both photographic portraits and artworks, specifically including Chinese ink-wash painting, colored pencil drawing, oil painting, and portrait photo. The generation results of READ under the same talking audio signal and these varied reference image conditions are illustrated in Fig.~\ref{fig:multipotraits}, where the frames presented in Fig.~\ref{fig:multipotraits} are sampled from identical timesteps across different generation results. The results demonstrate that for these reference styles unseen during training, READ achieves high-fidelity animations conditioned on the speech input, exhibiting both outstanding generalization performance and identity preservation capabilities. Notably, frames generated based on distinct reference images show a high degree of consistency in lip movements, aligned with the audio condition. These findings confirm the robust zero-shot generalization capability of READ across diverse reference image styles. Additional qualitative results are included in the supplementary video.

\subsection{Performance on Multiple Languages} As demonstrated in the supplementary video, our method generates satisfactory results for speech in multiple languages, such as French and Portuguese, despite being trained exclusively on English-only datasets HDTF~\cite{hdtf} and MEAD~\cite{mead}. This is primarily attributed to the robust cross-lingual generalization capabilities of the pre-trained Whisper-tiny encoder, which significantly enhances the READ framework’s adaptability in generating talking-head videos across diverse linguistic contexts, substantially improving the practicality of READ.

\section{Limitations and Future Work}
\label{sec:futurework}
Despite READ's promising advancements in expressive and efficient audio-driven talking head generation, it still encounters several challenges that open the way for future research. Firstly, although the READ framework generates realistic head movements, it occasionally produces motion blur artifacts during large-amplitude head movements, impacting the overall visual quality of the output video. This issue is primarily related to the quality of the training dataset, specifically the presence of motion blur in samples with large-amplitude head motions. To mitigate this, two key approaches can be employed: (1) refining the training dataset to exclude samples with excessive motion, and (2) applying a video motion blur detector or manual screening to identify and filter out videos containing significant motion blur.

Secondly, the READ framework occasionally exhibits insufficient clarity in the dental region, occasionally generates teeth that appear blurred or lack definition when the reference image provides an insufficient or occluded reference for the subject's teeth (e.g., a closed-mouth portrait). Future work could mitigate this issue through creating a high-fidelity training dataset of high-resolution talking head videos with well-defined teeth, and incorporating stronger facial priors as condition, such as those provided by external modules like ConsisID~\cite{consisid}, to offer explicit guidance for teeth reconstruction.

Despite these challenges, READ demonstrates significant potential for practical application, owing to its exceptional performance and remarkable inference speed. READ achieves an effective balance between generation quality and efficiency, making it highly suitable for real-time applications such as human-computer interaction and remote education. Consequently, READ holds substantial research and engineering value, establishing a solid foundation for future research in fast diffusion-based talking head generation.

\section{Ethical Consideration}
\label{sec:ethicalconsideration}
Our proposed READ framework can generate photorealistic talking head videos that are difficult to distinguish from genuine footage, suggesting a wide range of practical applications. These potential uses, spanning domains from human-computer interaction and remote education to caregiving companionship, offer significant societal and technological benefits. However, the same capabilities present a substantial risk of malicious use, particularly for the creation and dissemination of misinformation and harmful content. For example, the technology could be exploited to fabricate deceptive videos of public figures, produce violent or sexually explicit material, or create counterfeit media for purposes like extortion. Such misuses would yield severe negative consequences, undermining the fundamental goal of our research: to harness AI for the betterment of society.

Recognizing these risks, we have embedded ethical considerations and safety protocols throughout the development lifecycle of READ. During the training stage, we performed meticulous data filtering to rigorously exclude any content involving violence, sexual themes, or other inappropriate material. Furthermore, we have implemented strict usage controls on the deployment of READ. The current research-purpose deployment of READ is under our supervision of our risk assessment team. All inputs (images and audio) are subject to a stringent review to prevent the generation of malicious content. For any potential future releases of READ, we are committed to establishing a stringent review and assessment process to guarantee that generated content remains free of harmful materials. Moreover, we strongly advocate for research on advanced forgery detection techniques. The development of robust methods for identifying synthetic videos is crucial for the entire field of talking head video generation to collectively mitigate the risks of misuse. We are resolute in our commitment to addressing and preventing any potential misuse of READ.

\section{Summary} 
This supplementary PDF provides technical details supporting the main manuscript. Sec.~\ref{sec:dataset} introduces the detailed rationale for dataset selection along with the data pre-processing pipeline, providing a supplement to Sec.~\ref{sec:setup} of the main paper. Subsequently, Sec.~\ref{sec:architecture} presents the architectural details of our proposed READ framework, providing a supplement to Sec.~\ref{sec:framework}. Then, Sec.~\ref{sec:trainvaldetails} elaborates on the training and inference details of READ. In addition, in Sec.~\ref{sec:additionalresults} we provide an extensive qualitative evaluation of the performance and generation diversity under diverse speech and reference image conditions, thereby validating the robustness of the proposed framework. Furthermore, Sec.~\ref{sec:futurework} includes a discussion on the limitations of our model and potential research directions for future work. Finally, Sec.~\ref{sec:ethicalconsideration} discusses the ethical considerations related to our proposed technology. Collectively, this supplementary PDF delivers exhaustive details essential for reproducibility of READ while also demonstrating the significant advancements and value of our proposed framework in the research field of diffusion-based audio-driven talking head generation.

\end{document}